\documentclass[prc, showpacs, superscriptaddress, preprint, onecolumn]{revtex4}
\usepackage{graphics}
\usepackage{flushend}
\usepackage[bookmarksopen, bookmarksnumbered]{hyperref}
\newcommand{\nscl}{\affiliation{National Superconducting Cyclotron
    Laboratory and Department of Physics and Astronomy at Michigan
    State University, East Lansing, MI 48824, USA}}
\newcommand{\lpc}{\affiliation{LPC/Ensicaen and University of Caen, 6 Bd du Mar{\'e}chal Juin, F-14050 Caen cedex, France}}
\newcommand{\washu}{\affiliation{Chemistry Department, Washington University, St. Louis, MO 63130, USA}}
\newcommand{\uw}{\affiliation{Department of Physics, University of Wisconsin, Madison, WI 53706, USA}}
\newcommand{\gsi}{\affiliation{Gesellschaft f{\"u}r Schwerionenforschung mbH, D-64291 Darmstadt, Germany}}
\newcommand{\infn}{\affiliation{Laboratori Nazionali del Sud and INFN, I-95123, Catania, Italy}}
\newcommand{\nipne}{\affiliation{NIPNE, RO-76900 Bucharest-Magurele, Romania}}
\newcommand{\rio}{\affiliation{Instituto de Fisica, Universidade Federal do Rio de Janeiro, Cidade Universit{\'a}ria, CP 68528, 21945-970 Rio de Janeiro, Brazil}}
\newcommand{\tamu}{\affiliation{Cyclotron Institute, Texas A\&M University, College Station, TX 77843, USA}}

\begin{document}
\title{Comparisons of Statistical Multifragmentation and Evaporation
  Models for Heavy Ion Collisions}
%%%%%
\author{M.~B.~Tsang}\nscl 
\author{R.~Bougault}\lpc 
\author{R.~Charity}\washu
\author{D.~Durand}\lpc
\author{W.~A.~Friedman}\uw
\author{F.~Gulminelli}\lpc 
\author{A.~Le~F{\`e}vre}\gsi 
\author{Al.~H.~Raduta}\infn\nipne
\author{Ad.R.~Raduta}\nipne 
\author{S.~Souza}\rio 
\author{W.~Trautmann}\gsi
\author{R.~Wada}\tamu

\date{\today}

\begin{abstract}
The results from ten statistical multifragmentation models
have been compared with each other using selected experimental
observables. Even though details in any single observable may
differ, the general trends among models are similar. Thus these
models and similar ones are very good in providing important physics
insights especially for general properties of the primary fragments
and the multifragmentation process. Mean values and ratios of
observables are also less sensitive to individual differences in the
models. In addition to multifragmentation models, we have compared
results from five commonly used evaporation codes. The fluctuations
in isotope yield ratios are found to be a good indicator to evaluate
the sequential decay implementation in the code. The systems and the
observables studied here can be used as benchmarks for the development
of statistical multifragmentation models and evaporation codes.
\end{abstract}

\maketitle
%%%%%%%%%%%%%%%%%%%%%%%%%%%%%%%%%%%%%%%%%%%%%%%%%%%%%%%%%%%%%%%%%%%%%%%%%%%%%%%%%%%%%%%%%%%%%%%%%%%%%%%%%%%%
%% Body of the paper will go here:
%%%%%%%%%%%%%%%%%%%%%%%%%%%%%%%%%%%%%%%%%%%%%%%%%%%%%%%%%%%%%%%%%%%%%%%%%%%%%%%%%%%%%%%%%%%%%%%%%%%%%%%%%%%%
\section{Introduction}
%%%%%%%%%%%%%%%%%%%%%%%%%%%%%%%%%%%%%%%%%%%%%%%%%%%%
During the later stages of a central collision between heavy nuclei
at incident energies in excess of about $E/A$=50 MeV, a rapid
collective expansion of the combined system occurs \cite{ref1}.
Experimental evidence indicates that mixtures of intermediate mass
fragments (IMF's) with $3\leq Z\leq 30$ and light charged
particles (LCP, $Z\leq 2$) are emitted during this expansion stage.
With increased nucleon collisions, the properties of the nuclear
matter created can be described with equilibrium and statistical
concepts \cite{ref2,ref3,ref4,ref5,ref6,ref6b,ref6c,ref7,ref8,ref9}.
Ultimately, one would like to describe nuclear collisions with a
model that takes into account all the dynamics of nucleon-nucleon
collisions. Until then, statistical models provide invaluable
insight to the physics of multifragmentation of the last three
decades, by reducing the intractable problem of time-dependent
highly correlated interacting many-body fermion system to the much
simpler picture of a system of non-interacting clusters
\cite{ref10}.

Since most statistical multifragmentation codes have been developed
to describe specific sets of data and nearly all of them have
different assumptions, they are not equivalent
\cite{ref2,ref3,ref4,ref5,ref6,ref6b,ref6c,ref7,ref8,ref9,ref11,ref11b,ref12,ref13,ref14,ref15}.
One of the goals of this article is to examine the observables
constructed with the isotope yields from different statistical
multifragmentation models used in recent years. Even though the
number of models we studied is limited, they represent the codes
widely used in the heavy ion community. The results show that all
the statistical codes give similar general trends but different
predictions to specific experimental observables. The conclusion is
consistent with a recent study on models with different statistical
assumptions \cite{ref17}. We also find that the differences between
models are much reduced for observables constructed with isotope
yield ratios from different reactions.

The various codes and the bench mark systems which form the basis
for comparison will be described in Section 2. Comparisons of the
statistical multifragmentation models are presented in Section 3 and
the results from the comparisons of five different evaporation codes
are presented in Section 4. Finally we summarize our
 findings in Section 5.

%%% Table will go here:
\begin{table*}
\caption{Summary of the different statistical multifragmentation
models and evaporation codes studied in this article.}
\begin{center}
\begin{tabular}{|l|c|c|c|c|c|c|c|c|c|}
\hline
Code&Evaporation&user&author&Ref.&(168,75)&(186,75)&(168,50)&Primary&Final\\
\hline
\multicolumn{10}{|c|}{{\bf Statistical Multifragmentation Models}}\\
\hline ISMM-c&MSU-decay&Tsang&Das Gupta&2&Y&Y&&Y&Y\\\hline
ISMM-m&MSU-decay&Souza&Souza&13,14&Y&Y&&Y&Y\\\hline SMM95&own
code&Bougault&Botvina&4, 9&Y&Y&&Y&Y\\\hline MMM1&own code&AH
Raduta&AH Raduta&15&Y&Y&Y&Y&Y\\\hline MMM2&own code&AR Raduta&AR
Raduta&15&Y&Y&Y&Y&Y\\\hline MMMC&own code&Le F{\`e}vre&Gross&5,
16&Y&Y&Y&&Y\\\hline LGM&N/A&Regnard&Gulminelli&17&Y&&Y&Y&\\\hline
QSM&own code&Trautmann&St{\"o}cker&18&Y&Y&Y&&Y\\\hline
EES&EES&Friedman&Friedman&7, 8&Y&Y&Y&Y&Y\\\hline
BNV-box&N/A&Colonna&Colonna&24&Y&Y&&Y&\\
\hline
\multicolumn{10}{|c|}{{\bf Evaporation codes}}\\
\hline Gemini&&Charity&Charity&25&Y&Y&&&Y\\\hline
Gemini-w&&Wada&Wada&25--28&Y&Y&&&Y\\\hline
SIMON&&Durand&Durand&29&Y&Y&&&Y\\\hline EES&&Friedman&Friedman&7,
8&Y&Y&Y&&Y\\\hline
MSU-decay&&Tsang&Tan \textit{et al.}&14&Y&Y&&&Y\\
\hline
\end{tabular}
\end{center}
\end{table*}

%%%%%%%%%%%%%%%%%%%%%%%%%%%%%%%%%%%%%%%%%%%%%%%%%%%%%%%%%%%%
\section{Bench-mark systems}
%%%%%%%%%%%%%%%%%%%%%%%%%%%%%%%%%%%%%%%%%%%%%%%%%%%%%%%%%%%%%
Nearly all statistical models assume that nucleons and fragments
originate from a single emission source characterized by $A_{0}$
nucleons and $Z_{0}$ protons. The hot fragments then de-excite using
evaporation models. To provide consistent comparisons between
models, we have chosen the following source systems: 1)
$A_{0}$=168, $Z_{0}$=75, $N_{0}$/$Z_{0}$=1.24, 2) $A_{0}$=186,
$Z_{0}$=75, $N_{0}$/$Z_{0}$=1.48. These two systems have the same
charge and are chosen to be 75\% of the initial compound systems of
$^{112}$Sn+$^{112}$Sn and $^{124}$Sn+$^{124}$Sn \cite{ref18,ref19}. We
also have calculations on system
 3) $A_{0}$=168, $Z_{0}$=84, $N_{0}$/$Z_{0}$=1.0 which
 has the same mass but different charge from system 1.
 Even though most results of system 3 are not included in this article due to lack of space, they corroborate
 the conclusions. In each calculation, the same inputs are used. We require the source excitation energy,
 $E^{\star}$, to be 5 MeV per nucleon
and the source density to be 1/6 of the normal nuclear matter
density.

At the time when this manuscript was prepared, we were able to get
results from nine statistical multifragmentation model codes plus a
hybrid dynamical-statistical code, (BNV-box) and five evaporation
codes. Table 1 lists all the codes, users (defined as the person who
did the calculations shown in this paper) and the main authors of
the codes. The users sent us the output files which contain mainly
the neutron ($N$) and proton ($Z$) number and the yield of the hot
fragments and/or the final fragments. All these output files can be
found in the web: http://groups.nscl.msu.edu/smodels/results.html.

The statistical multifragmentation models studied here construct
fragment yields from a maximum entropy principle, but they differ
both in the degrees of freedom employed and in the chosen
constraints. We have different versions of the Statistical
Multifragmentation Model 
\linebreak[4] 
(SMM) \cite{ref20}. All these models assume
that the N-body source correlations are exhausted by clusterization
and, therefore, describe the system as a collection of
% line break with second dash in non-inter-acting
non-inter-acting clusters. (The Coulomb repulsion among fragments
are approximately taken into account.) These codes differ in the
freeze out volume prescription, in the treatment of continuum states
and in the numerical technique to span the phase space. The SMM95
code uses grand-canonical approximation \cite{ref4,ref7} and
Fermi-jet breakups for the deexcitation of hot fragments. The
Improved Statistical Multifragmentation Model (ISMM)
\cite{ref11b} uses experimental masses and level densities when
available. When experimental information is not available, ISMM uses
an improved algorithm to interpolate level densities for the hot
fragments. It uses the MSU-decay code as an afterburner. ISMM-c
\cite{ref2} uses a canonical formalism, while 
\linebreak[4] 
ISMM-m \cite{ref11}
adopts a microcanonical approach. The sequential decay algorithm in
ISMM \cite{ref11} uses experimental masses and includes structure
information for light fragments $(Z<15)$. The MMMC code uses a
Metropolis-Monte Carlo method \cite{ref5,ref13}. MMMC is the only
model that can accommodate non-spherical sources but only neutrons
are emitted in the sequential decays. We have two calculations using
the microcanonical multifragmentation model MMM \cite{ref12} with
different freeze-out assumptions: (1) non-overlapping spherical
fragments inside a spherical source and (2) free volume approach.
These two calculations are correspondingly denoted by MMM1 and MMM2.
The Quantum Statistical Model (QSM) \cite{ref15} is a simplified
grand-canonical version of SMM models including only a limited
number of light clusters $(A<20)$, which however are described with
a detailed density of states accounting for all known discrete
levels at the time when the code was written in the late eighties.

The Expanding Emitting Source (EES) model \cite{ref6b,ref6c} is an
extended Weisskopf evaporation model \cite{ref15b} which couples the
emission of fragments to the changing conditions, i.e., density
(volume), mass-number, isospin, and entropy, of the source. The
model assumes an equation-of-state for the source so that the
thermal pressure and initial expansion determine the changes in the
source due to emission.  It is the only statistical model to account
for the time dependence of the emission process. No specific density
(volume) for emission is assumed, but the model predicts that the
strongest emission often occurs from a dilute source during a narrow
time period.  (In this sense it is similar to the SMM.)  Spectra are
constructed by summing the contributions of emission from different
times, with a switch from surface to volume emission at a low
density of the source.

Finally, we have two microscopic model calculations. The Lattice Gas
Model (LGM) \cite{ref14} calculates the equilibrium configurations
of a system of (semi)classical nucleons interacting via an Ising
Hamiltonian. These configurations are generated in a given confining
box by Monte Carlo. It is the only model that has no nuclear physics
input. The BNV-box model is based on the Boltzmann-Nordheim Vlaslov
(BNV) equations \cite{ref16} and uses the effective Skyrme force
augmented with a stochastic collision integral to calculate the
equilibrium configurations which are generated via a dissipation
dynamics in a box. In both models, the clusters have to be defined a
posteriori via a clusterization algorithm.

Since de-excitation of the hot fragments is essential before
comparison  to experimental data, most codes have their own
sequential decay algorithms. Ideally, one should compare the hot
primary fragments and the decay fragments separately. Unfortunately,
in some codes (e.g. \linebreak[4]
 MMMC), the hot fragments cannot be extracted
while in others (LGM and BNV-box) the hot fragments sent by the
users have not undergone decay. This makes comparing the
contributions from the evaporation portion of the code to the final
fragments very difficult.

Since an "after-burner" or evaporation code is needed to allow
the hot  fragments to decay to ground states, codes that can be
coupled to statistical and dynamical codes are very important. Thus
in addition to the fragmentation models, we also compare five
different evaporation codes (listed in Table 1) that have served the
functions of "after-burners" to both statistical and dynamical
codes. 1.) The most widely used code is Gemini \cite{ref21} which
treats the physics of excited heavy residues very well. However, for
the light fragments, it lacks complete structural information. 2.) A
modified code of an early version of Gemini \cite{ref22,ref23} has
also been used extensively to de-excite hot fragments generated in
the Asymmetrized Molecular Dynamical (AMD) Model \cite{ref24}. We
labeled this modified version of Gemini as Gemini-w. 3.) An event
generator code called SIMON \cite{ref25}, based on Weisskopf
emission rates \cite{ref15b}, includes the narrowest discrete states
for $Z \leq 9$ as well as in-flight evaporation. It has been used to
deexcite fragments created in both BNV dynamical model \cite{ref26}
and a heavy-ion phase space model \cite{ref27}. 4.) The MSU-decay
code \cite{ref11b} uses the Gemini code to decay heavy residues and
includes much structural information such as the experimental masses,
excited states with measured spin and parity for light fragments
with $Z<15$ in a table. This table also includes information of
calculated states, which are not measured. 5.) In principle, at very
low excitation energy, the multifragmentation models can also be
used as evaporation models. In this category, we have results from
the EES model \cite{ref6b,ref6c}.

For the evaporation model comparison, the bench mark systems for the
source are the same as the three systems used in the
multifragmentation models, $(A_{0},Z_{0})=(168,75), (186, 75)$ and
$(168, 84)$. The excitation energy is set to be 2 MeV per nucleon
and the density is assumed to be the same as normal nuclear matter
density.

%%%%%%%%%%%%%%%%%%%%%%%%%%%%%%%%%%%%%%%%%%%%%%%%%%%%%%%%%%%%%%%%%%%%%%%%%%%%%%%%%%%%%%%%%%%
%% Section 3
\section{Results from multifragmentation models}
%%%%%%%%%%%%%%%%%%%%%%%%%%%%%%%%%%%%%%%%%%%%%%%%%%%%%%%%%%%%%%%%%%%%%%%%%%%%%%%%%%%%%%%%%%%%
In this section, we choose to show results that illustrate the
differences  and similarities between calculations. Due to limited
space, not all observables from the calculations are constructed or
shown here. Since system 1 with $A_{0}$=168 and $Z_{0}$=75 have
results from all the calculations, we tend to highlight this system.
Some of the results on the ISMM-c calculations have been published
in ref.~\cite{ref2}. If a choice has to be made between showing
ISMM-c results or ISMM-m results due to lack of space, we choose to
show the results of ISMM-m. For the LGM calculations
\cite{ref14,ref28}, we have results using the micro-canonical
approximations as well as results using canonical approximations. We
show mainly the results with the microcanonical approximations. The
differences between the microcanonical and canonical assumptions can
be inferred from the results of ISMM-c and ISMM-m. The observables
shown in Sections 3.1 to 3.5 are chosen for the relevance of the
observables to the understanding of the multifragmentation process.
More recently, the focus of heavy ion collisions at intermediate
energy has shifted to explore the isospin degree of freedom
\cite{ref2}. This is often done by studying two or more systems,
which differ mainly in the isospin composition of the projectiles or
targets. Isoscaling using isotope yield ratios is discussed in
Section 3.4. Instead of using isotope yield ratios for temperature,
we use the fluctuations of different thermometers to determine how
well the sequential decays in the code reproduce the observed
fluctuations. The results will be described in Section 3.5.

To provide some uniformity to the figures, we will try to use the
same  symbols for the results from the same code throughout this
article. Where applicable, closed symbols often refer to the neutron
rich system ($A_{0}$=186, $Z_{0}$=75) and open symbols refer to the
neutron deficient system ($A_{0}$=168, $Z_{0}$=75). We also adopt
the convention that the results are labeled with the user (who sent
us the calculated results) and the code name. Even though comparison
with data is not our goal, it is sometimes instructive to plot the
data as reference points when appropriate. We have chosen the data
from the central collisions of $^{124}$Sn+$^{124}$Sn and
$^{112}$Sn+$^{112}$Sn \cite{ref18,ref19} at 50 MeV per nucleon
incident energy as represented by closed and open star symbols,
respectively, mainly because this data set is readily available to
the first author.
%%%%%%%%%%%%%%%%%%%%%%%%%%%%%%%%%%%%%%%%%%%%%%%%%%%%%
\subsection{IMF multiplicities}  %% 3.1
%%%%%%%%%%%%%%%%%%%%%%%%%%%%%%%%%%%%%%%%%%%%%%%%%%%%%
The copious production of intermediate mass fragments (IMF) which
are  charged particles with $Z$=3-20 is one signature of the
multifragmentation process. The study of these fragments provides clues
to the nuclear liquid-gas phase transition as they are considered as
droplets formed from the condensation of nuclear gas and may provide
information about the co-existence region.  Figure 1 shows the mean
multiplicities of IMFs produced by different models. Within errors,
one cannot discern any dependence of the mean IMF multiplicity on
the isospin composition of the initial sources by looking for
systematic differences between solid and open symbols which
represent the neutron-rich and neutron-deficient systems,
respectively. If we compare the left and right panel of Fig. 1, in
general, sequential decays reduce the IMF multiplicities.

\begin{figure}
\centering
\resizebox{1.0\columnwidth}{!}{%
\includegraphics{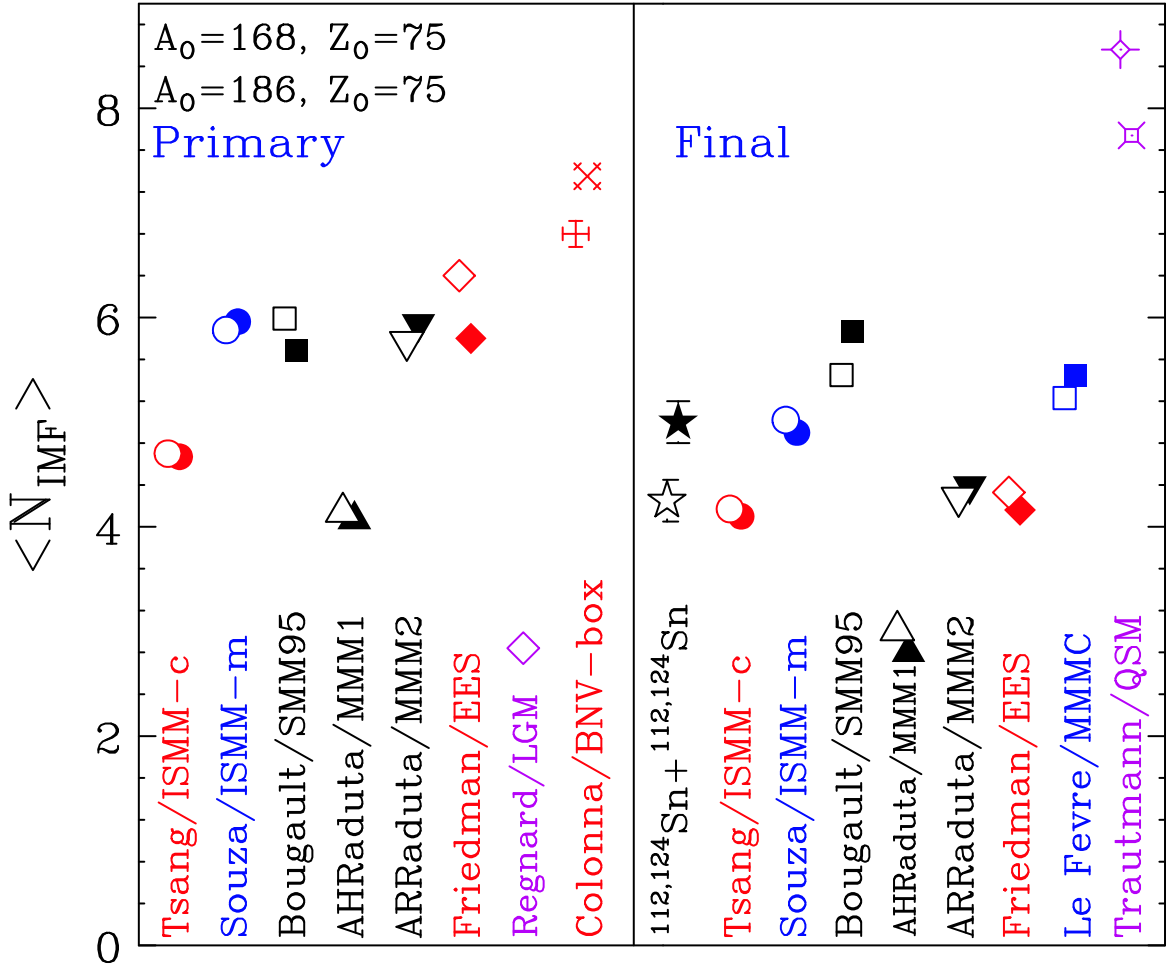}}
\caption{Mean IMF multiplicity obtained from different statistical
models listed in Table 1 for primary fragments (left panel) and
final fragments (right panel). At the bottom of the panels, the
calculations are labeled by the name of the user and the name of the
code. The open symbols refer to system 1 ($A_{0}$=168, $Z_{0}$=75)
and the solid symbols refer to system 2 ($A_{0}$=182, $Z_{0}$=75).
The open and solid stars in the right panel are data from the
central collisions of $^{112}$Sn+$^{112}$Sn and
$^{124}$Sn+$^{124}$Sn systems at $E/A$=50 MeV \cite{ref32}.}
\label{fig1}
\end{figure}

The two MMM calculations have different results due to different
freeze  out assumptions used for the source. MMM1 which uses
non-overlapping spherical fragments emits nearly two fragments
less than MMM2. Only primary fragments before decay are available
from the LGM and BNV-box calculations. For the MMMC and QSM models,
we only have fragments after decay.

For the primary fragment multiplicity (left panel), the BNV model
emits  slightly more primary fragments while the LGM model emits
nearly a factor of two less fragments than the other models. For the
final fragment multiplicity (right panel), the QSM
\cite{ref15,ref29,ref30} emits many more IMFs.  Indeed this model is
not suited to predict absolute yields but rather should be used to
compute relative yields of light isotopes e.g. for thermometry
purposes \cite{ref30,ref31}. For comparisons, the data from the
central collisions of Sn isotopes are represented by the star
symbols in the right panel. The differences in the mean
multiplicities between the $^{112}$Sn+$^{112}$Sn (open stars) and
$^{124}$Sn+$^{124}$Sn (solid stars) \cite{ref32} are much larger
than those predicted by all the models after decay. The
discrepancies between model predictions and data are not understood.

%%%%%%%%%%%%%%%%%%%%%%%%%%%%%%%%%%%%%%%%%%%%%%%%%%%%%%%
\subsection{Mass Distributions}   % 3.2
%%%%%%%%%%%%%%%%%%%%%%%%%%%%%%%%%%%%%%%%%%%%%%%%%%%%%%%

\begin{figure}
\centering
\resizebox{1.0\columnwidth}{!}{%
\includegraphics{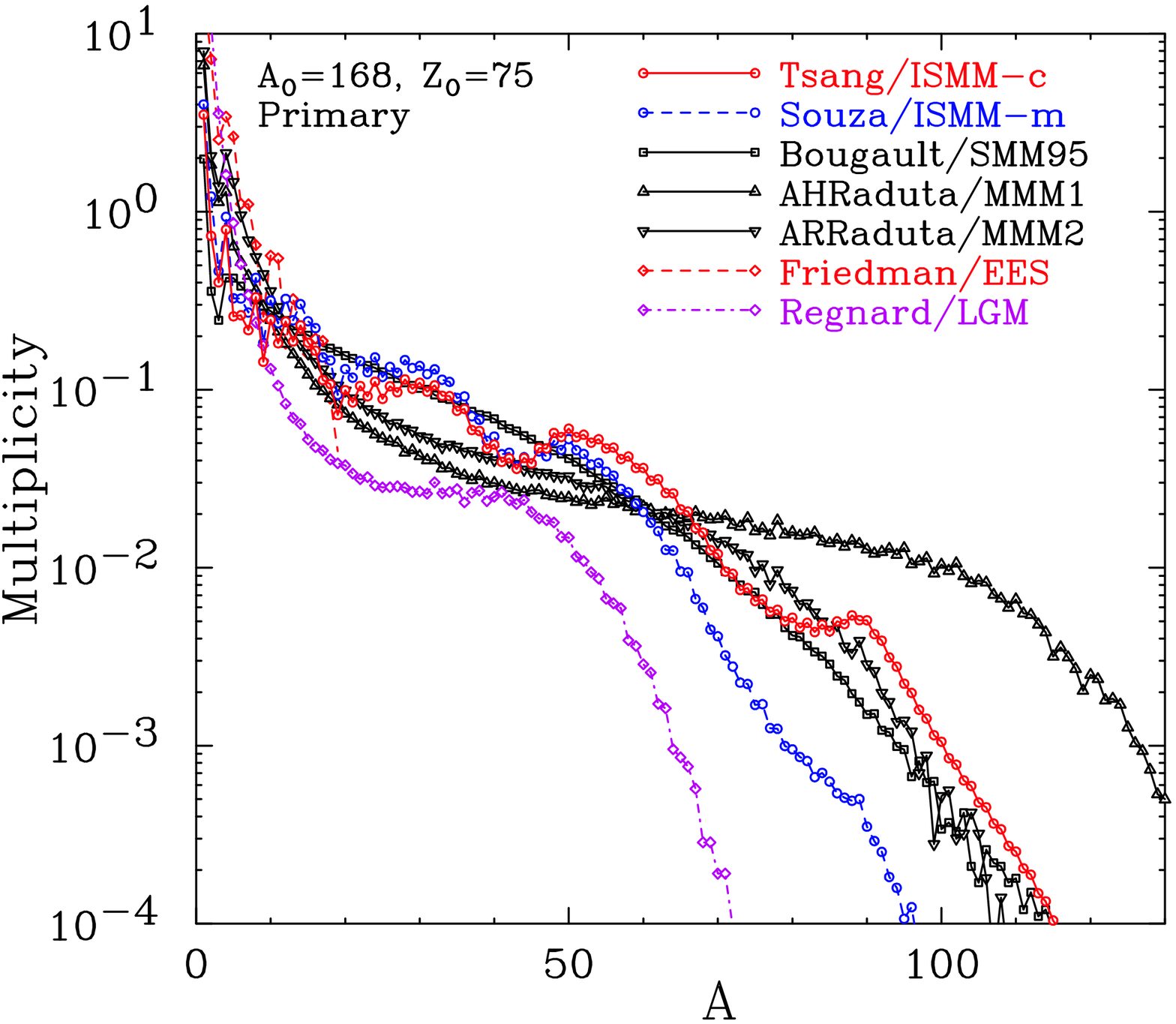}}
\caption{Predicted primary fragment mass distributions from the
mutifragmentation of a source nucleus with $A_{0}$=168,
$Z_{0}$=75 (system 1).
See caption of Fig. 1 for name convention.
%For easier identification, please note that the curves cross the multiplicity level of 10$^{-3}$ in the sequence LGM, ISMM-m, SMM95, MMM2, ISMM-c, MMM1 (from left to right); the EES distribution extends to $A$=20 only.
}
\label{fig2}
\end{figure}

Next we examine the primary mass distributions of the $A_{0}$=168,
$Z_{0}$=75 system. The steep drop of the light fragment ($A<10$)
multiplicity shown in Fig.~2 are similar for nearly all the models
but there are differences. Some of the differences (e.g. between
MMM1 (upright triangles) and MMM2 (inverted triangles) arise from
differences in the freeze-out assumptions as described previously.
The differences in the results from the two ISMM calculations may
come from the difference between canonical and micro-canonical
approximations used. (ISMM-c requires temperature instead of
excitation energy as one property of the initial source.) SMM95 and
the two MMM calculations have smooth distributions as the fragment
masses are determined from liquid drop mass-formulae \cite{ref33}.
The LGM (the lowest curve with diamond symbols), which does not take
into account the binding energies or nucleon masses shows a smooth
dependence on mass but does not produce heavy residues.

\begin{figure}
\centering
\resizebox{1.0\columnwidth}{!}{%
\includegraphics{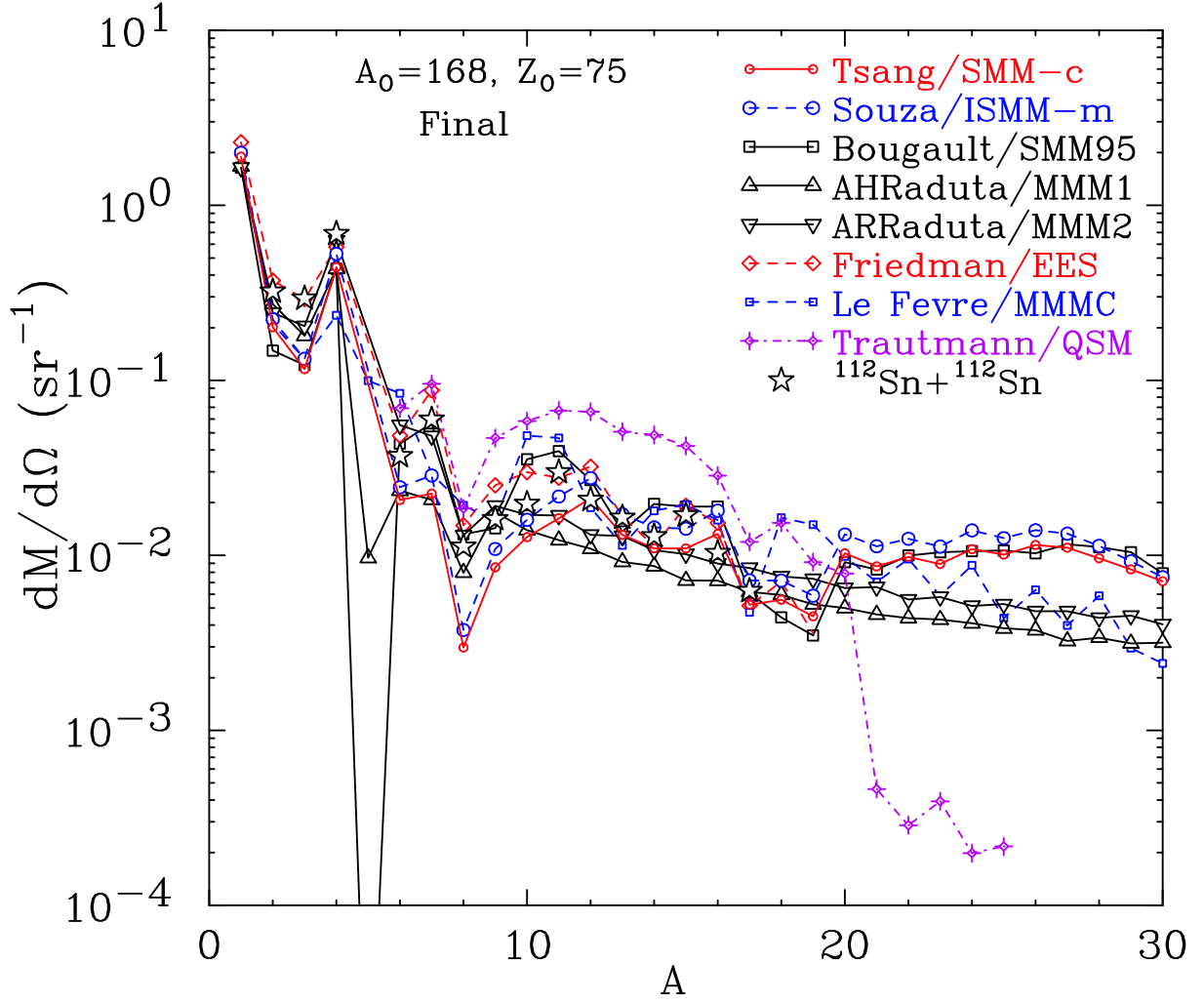}}
\caption{Predicted final fragment mass distributions from the
multifragmentation of a source nucleus with $A_{0}$=168, $Z_{0}$=50
(system 1). For comparison, data from the multifragmentation of
central collisions of $^{112}$Sn+$^{112}$Sn system \cite{ref2,ref18}
at $E/A$=50 MeV are plotted as open stars.} \label{fig3}
\end{figure}

In Figure 3, we have plotted the differential multiplicity of the
final mass distributions for the same  ($A_{0}$=168, $Z_{0}$=75)
system in an expanded scale. Again, while the trends are similar for
most calculations except the QSM model (crosses), there are
significant differences in detailed comparisons.  Most models do not
have nuclei with mass 5 and cross-sections for mass 8 are much
reduced in accordance to experimental observation. For reference,
the data from the $^{112}$Sn+$^{112}$Sn system are plotted as open
stars. The trends exhibited by most models are similar to those of
the data. Primary and final fragments with $A\geq20$ are ignored
in the EES code. These heavy fragments are not included in the
output files. The QSM does not produce fragments with $A>20$. To
conserve the total number of nucleons, more light charged fragments
with $A\leq20$ are produced, causing the over-production of IMFs
seen in both Figs.~1 and 3.

The charge distributions are similar to the mass distributions so
they are not discussed here.

%%%%%%%%%%%%%%%%%%%%%%%%%%%%%%%%%%%%%%%%%%%%%%%%%%%%%%%%%%
\subsection{Isospin observables and isotope distributions}   %% 3.3
%%%%%%%%%%%%%%%%%%%%%%%%%%%%%%%%%%%%%%%%%%%%%%%%%%%%%%%%%

One observable to study the isospin degrees of freedom is the
asymmetry, $N/Z$,  of the fragments. Figure 4 shows $<N/Z>$ as a
function of the fragment charge number $Z$ predicted by different
models. In this plot, the left panel shows results from the
neutron-deficient system ($A_{0}$=168, $Z_{0}$=75) while the right
panel contains results from the neutron-rich system ($A_{0}$=186,
$Z_{0}$=75). Unlike the mass distributions shown in Figs.~2 and 3,
differences between different models are not very large, about 10\%.
(The zero of the vertical axis is suppressed in order to show the
differences in more details.) As expected, the $<N/Z>$ of the
fragments are larger for the more neutron rich system. However, the
$<N/Z>$ values are much lower than the $<N_{0}/Z_{0}>$ of the
initial system of 1.48 for the neutron-rich system. For the
neutron-deficient system in the left panel, the initial
$<N_{0}/Z_{0}>$ value is 1.24 which is only slightly larger than the
fragment values.  For reference, data from the central collisions of
$^{124}$Sn+$^{124}$Sn (solid stars) and
\linebreak[4]
$^{112}$Sn+$^{112}$Sn systems
(open stars) \cite{ref18} are plotted in the left and right panels,
respectively. Since the excited fragments in MMMC only emit neutrons
\cite{ref13}, the fragment $<N/Z>$ (squares) are lower than those
derived from other models. All the other calculations exhibit
similar trends as the data.

\begin{figure}
\centering
\resizebox{1.0\columnwidth}{!}{%
\includegraphics{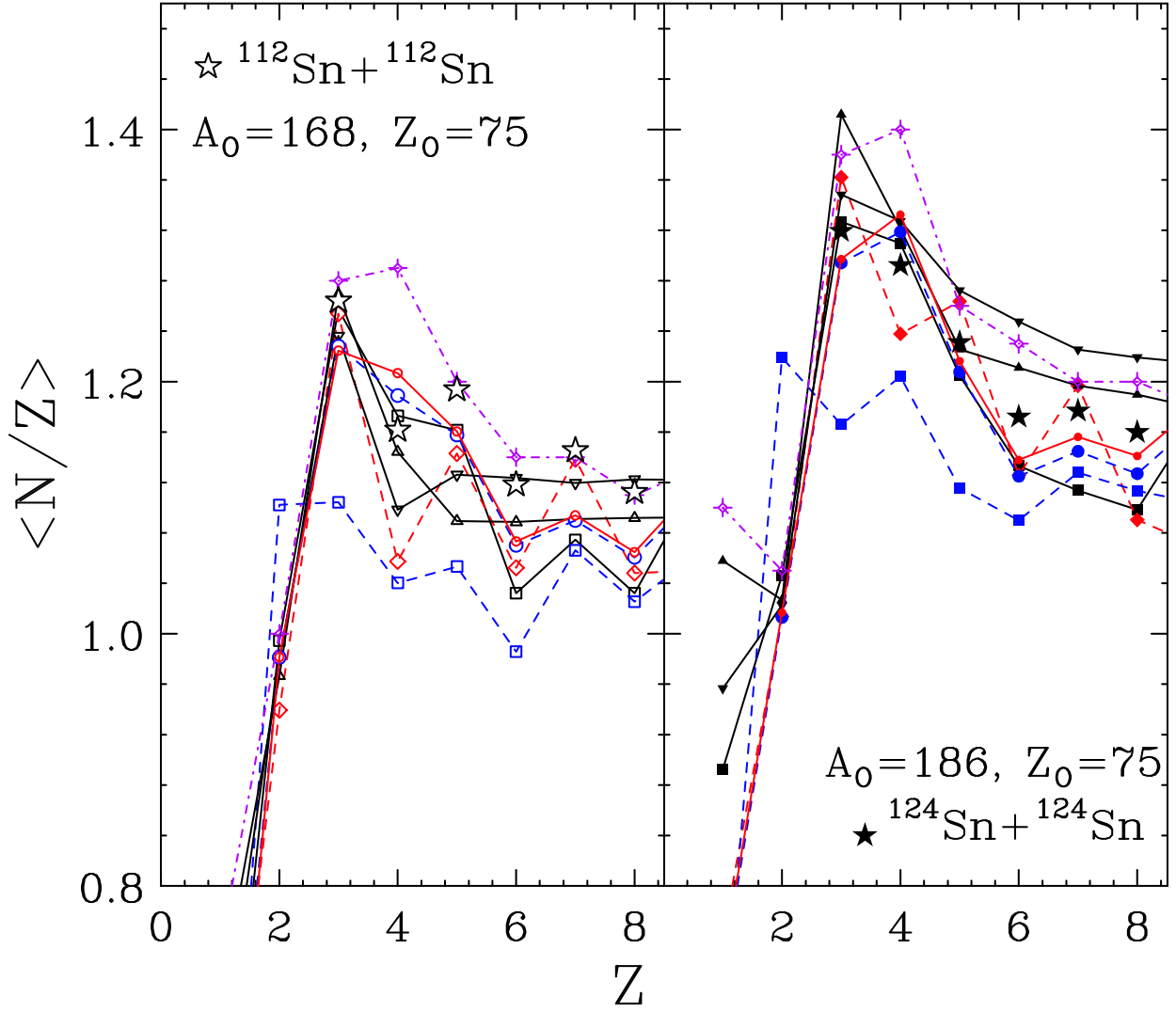}}
\caption{The mean neutron to proton ratios as a function of the
charge  of the emitted fragment $Z$ for system 1 (left panel) and
system 2 (right panel). For comparison, results from the
multifragmentation following central collisions of $^{112}$Sn+$^{112}$Sn
and $^{124}$Sn+$^{124}$Sn are shown as open (left panel) and closed stars
(right panel), data from ref.~\cite{ref18}.}
\label{fig4}
\end{figure}

As the average values of the asymmetry of the fragments are
determined from  the isotope yields, it is instructive to examine
the isotope distributions directly. Figure 5 shows the oxygen
isotope distributions from different models before (left panels) and
after (right panels) sequential decays. The upper panels indicate
the isotopes from the neutron-rich ($A_{0}$=186, $Z_{0}$=75) system
while fragments from the neutron-deficient ($A_{0}$=168, $Z_{0}$=75)
system are plotted in the bottom panels. For reference, data
\cite{ref18} from the central collisions of $^{124}$Sn+$^{124}$Sn
(solid stars) and
\linebreak[4]
$^{112}$Sn+$^{112}$Sn systems (open stars) are
plotted in the upper right panel and lower right panels,
respectively.

The differences in the primary distributions between models (left
panel)  can be understood from the nuclear masses used in the
different codes. Both the ISMM models (circle symbols) used
experimental masses even for hot fragments \cite{ref2,ref11b}, thus
odd-even effects are evident in the primary mass distributions. The
SMM95 (squares) \cite{ref7} and the two MMM (upright and inverted
triangles) \cite{ref12,ref33} calculations use mass formulae
resulting in smooth interpolations of isotope cross-sections. The
deficiency of models like the LGM (open diamond symbols in the lower
left panel), which do not include any nuclear physics information,
is obvious. The EES results are not presented here as the model
ignores primary and secondary fragments with $A/geq$20 and the
oxygen isotope yields are not complete.

The isotope distributions from all models after decay (right panels)
become  much narrower and resemble that of the experimental data.
The ISMM-c, ISMM-m and SMM95 models predict a peak at $^{16}$O due
to its large binding energy and the use of experimental masses in
the decays. The ISMM calculations that incorporate the MSU-decay
algorithms with experimental masses and structural information
exhibit odd-even effects. In the decay code of the MMM calculations,
fragment masses are derived from mass formulae \cite{ref33}. As a
result, the isotope distributions are rather smooth. The individual yields
of oxygen
isotopes are not available from the QSM output files, and the results
of this model is not represented here.

\begin{figure}
\centering
\resizebox{1.0\columnwidth}{!}{%
\includegraphics{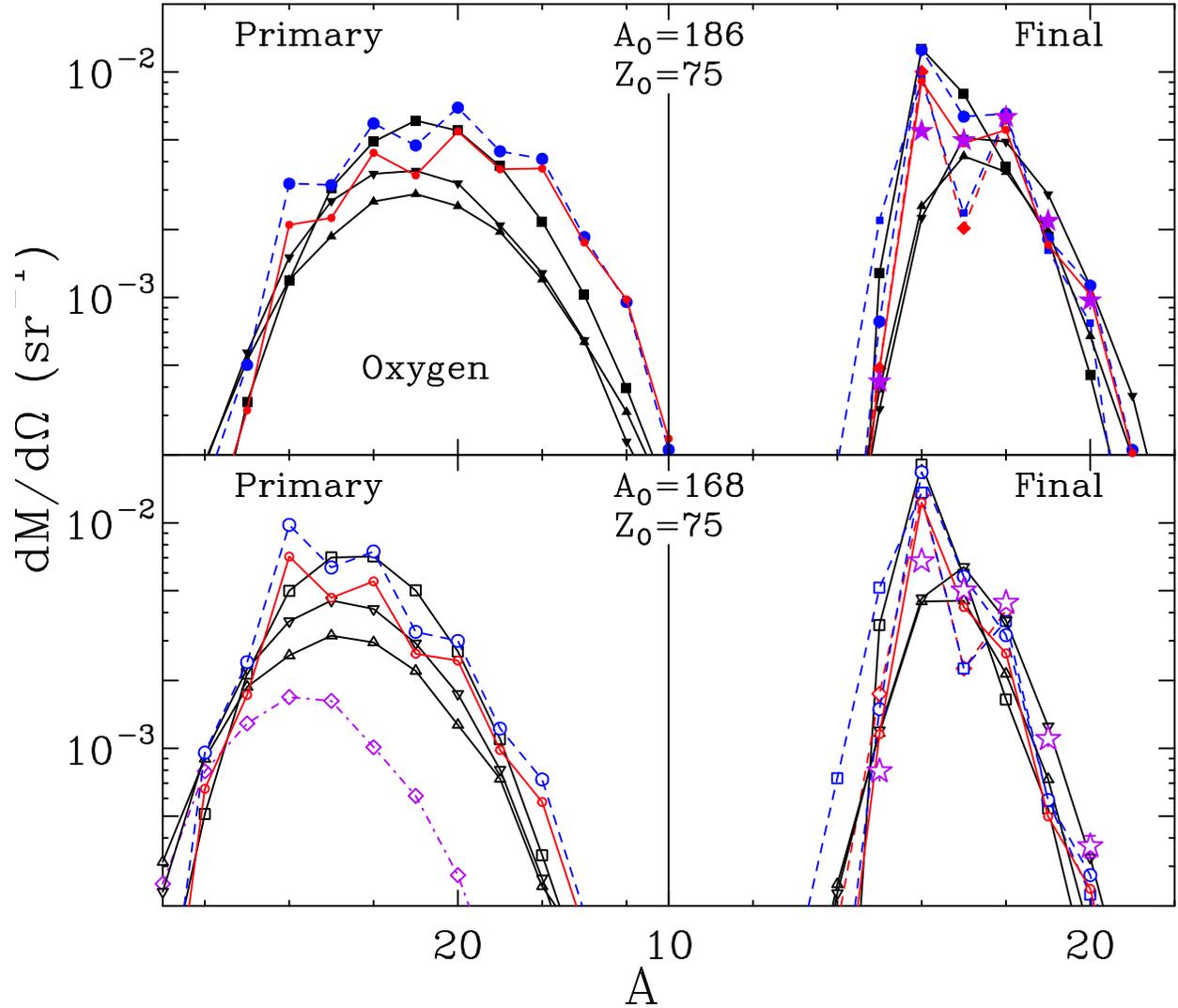}}
\caption{Predicted isotope distributions for oxygen fragments from
different  models. Primary fragments are plotted in the left panels
and final fragments in the right panels. The top panels contain
results from the neutron-rich system 2 and the bottom panels contain
distributions from the neutron-deficient system 1.  The open (bottom
right panel) and solid stars (top right panel) are data from
ref.~\cite{ref18}.} \label{fig5}
\end{figure}

In order to quantify the mean and the width of the distributions, we
have plotted the mean  mass number and the standard deviations of
the oxygen distributions in Fig.~6 for both the primary (left
panel) and final (right panel) fragment distributions. The vertical
bars represent the standard deviations of the isotope distributions.
In general, all models produce much wider distributions for the
primary isotopes and the widths are reduced by sequential decay
effects. Sequential decays tend to move the centroids of the
distributions towards the valley of stability and reduce the
differences in the centroids of the isotope distributions between
the neutron-rich and neutron-deficient systems.

%%%%%%%%%%%%%%%%%%%%%%%%%%%%%%%%%%%%%%%%%%%%%%%%%%%%%%%%
\subsection{Isoscaling}           % 3.4
%%%%%%%%%%%%%%%%%%%%%%%%%%%%%%%%%%%%%%%%%%%%%%%%%%%%%%%%%%%%%%%
When isoscaling was first observed in experimental data
\cite{ref19,ref34}, it  was demonstrated through statistical model
calculations that isoscaling could be preserved through sequential
decays \cite{ref34,ref35}. More importantly, statistical models
relate the isoscaling phenomenon to the symmetry energy
\cite{ref34,ref35,ref36,ref37}, which is of fundamental interest to
general nuclear properties as well as astrophysical interests
\cite{colonnabook}.

\begin{figure}
\centering
\resizebox{1.0\columnwidth}{!}{%
\includegraphics{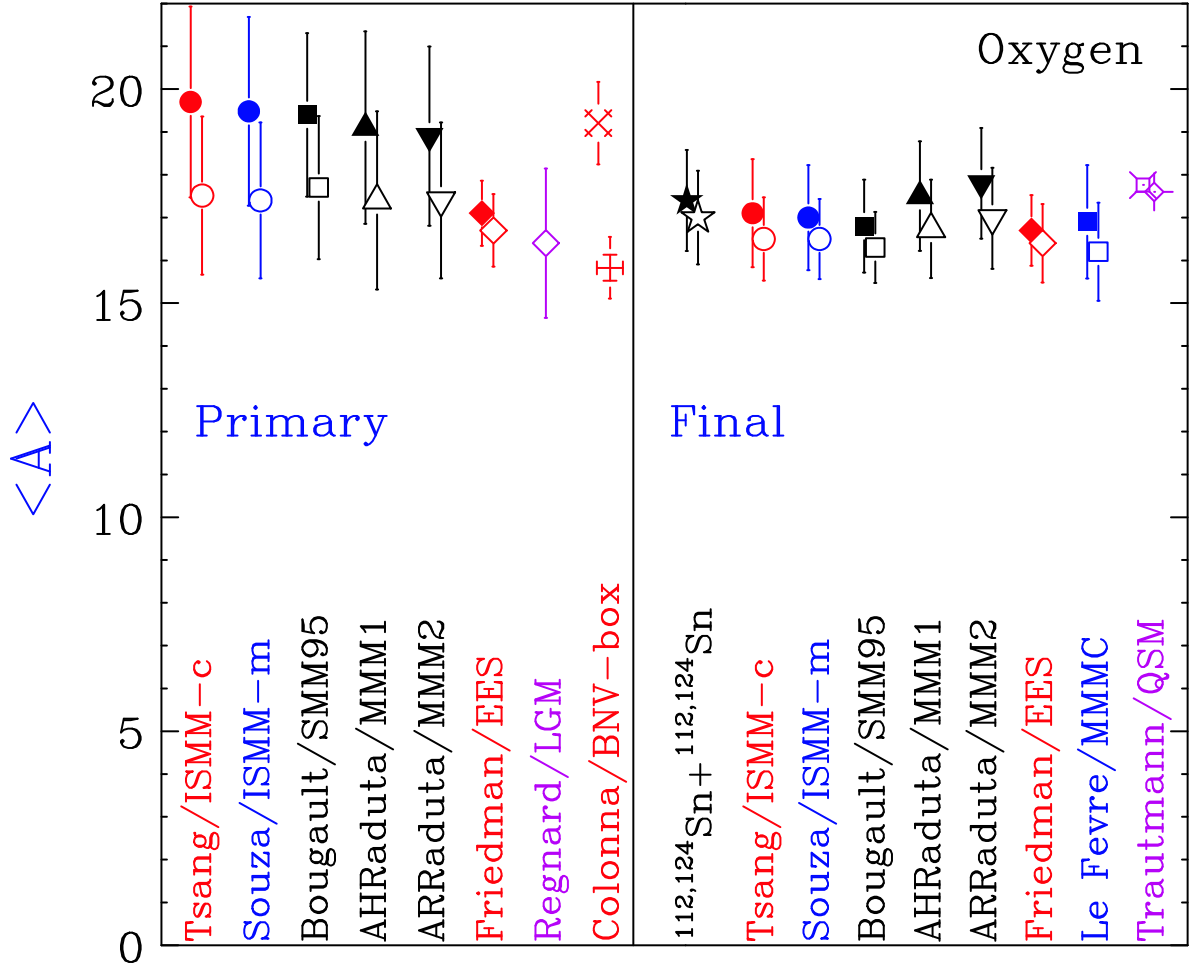}}
\caption{Centroids and widths (variance) of the oxygen isotope
distributions  obtained from different models. Most of the
distributions are shown in Fig.~5.} \label{fig6}
\end{figure}

Isoscaling describes the exponential dependence on the isotope neutron
($N$) and proton ($Z$) number of the yield ratios from two different
reactions,
\begin{equation}
R_{21}=\frac{Y_2(N,Z)}{Y_1(N,Z)}=C\mathrm{e}^{\alpha N + \beta Z}
\end{equation}
where $C$, $\alpha$, and $\beta$ are the fitting parameters. In our
specific examples of systems 1 and 2, $Y_2(N,Z)$ is the isotope
yield emitted from the neutron-rich system $A_{2}$=186, $Z_{2}$=75
and $Y_1(N,Z)$ is the isotope yield emitted from the neutron-deficient
system $A_{1}$=168, $Z_{1}$=75. Figure 7 shows that all statistical
multifragmentation models exhibit good isoscaling behavior for the
primary fragments. Each symbol corresponds to one element, $Z$=1 (open
triangles), $Z$=2 (closed triangles), $Z$=3 (open circles), $Z$=4 (closed
circles), $Z$=5 (open squares), $Z$=6 (closed squares), $Z$=7 (open
diamonds) and $Z$=8 (closed diamonds). The solid and dashed lines are
best fits from Eq.~(1). The slopes of the lines correspond to the
neutron isoscaling parameter $\alpha$ and the distance between the
lines corresponds to the isoscaling parameter $\beta$. All the
models except LGM have similar slopes. The slope parameters from
the two MMM models are slightly smaller. The LGM only has
calculations on systems 1 and 3. Since the differences in the
asymmetries between systems 1 and 2 and systems 1 and 3 are small,
the LGM isoscaling slopes are expected to be slightly smaller but
they are much smaller (lower left panel) than the other models. This
is probably related to the lack of nuclear mass input in such model.

\begin{figure}
\centering
\resizebox{1.0\columnwidth}{!}{%
\includegraphics{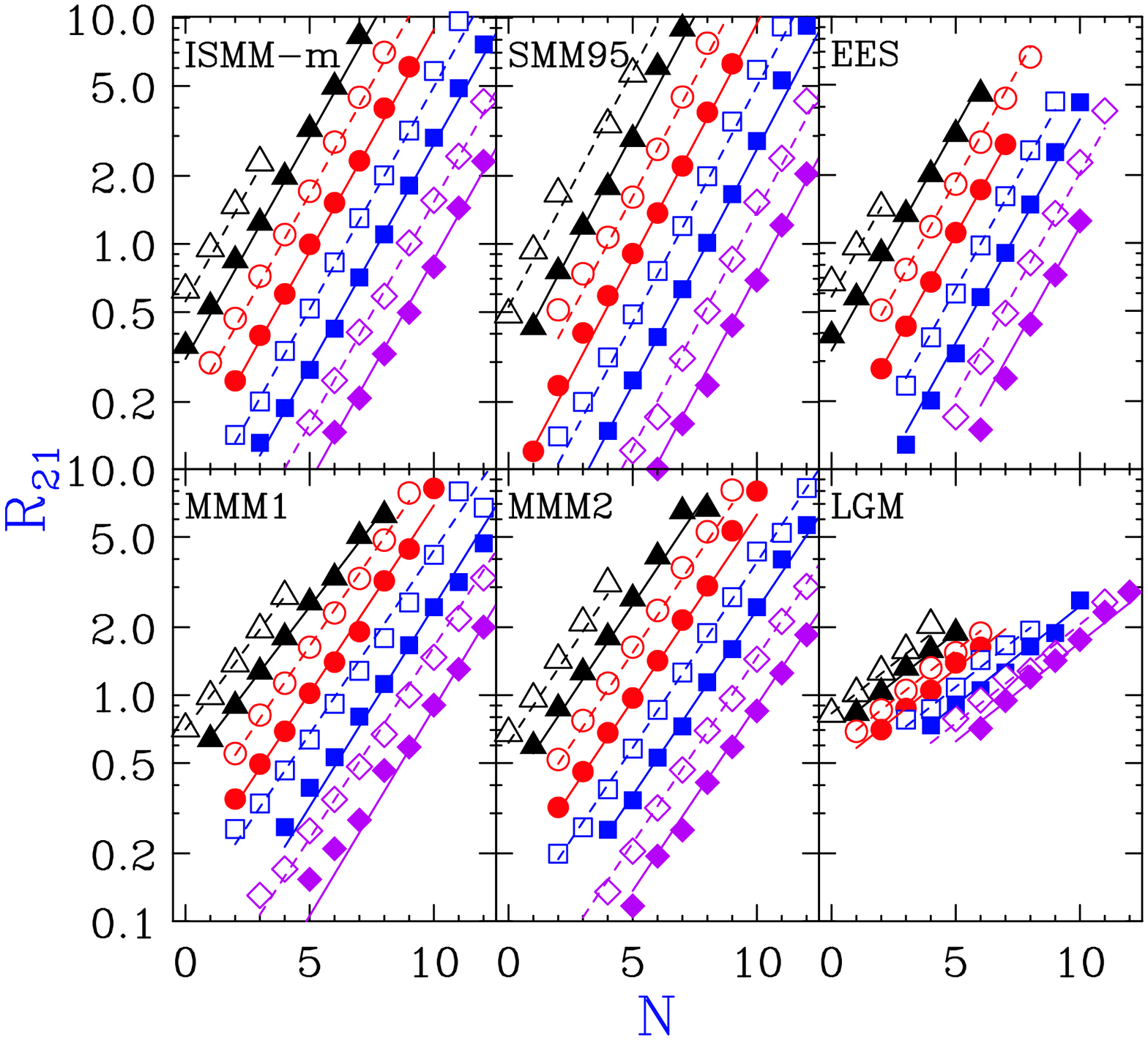}}
\caption{Predicted yield ratios, $R_{21}(N,Z) = Y_{2}(N,
Z)/Y_{1}(N, Z)$ from primary fragments for system 2 and system 1.
(For the LGM, the calculations are for system 1 and system 3.) Each
panel presents the results from one model calculation. The lines are
best fits to the symbols according to Eq.~(1).
 Different lines correspond to atomic numbers $Z$=1 to 8 starting with the leftmost
 line being $Z$=1. Open points and dashed lines denote
 isotopes with odd $Z$ while solid points and solid lines denote isotopes with even $Z$.
} \label{fig7}
\end{figure}

An important contribution that statistical models make to the field
of heavy ion collision is the derivation that the isocaling
parameter $\alpha$ is related to the symmetry energy coefficient,
$C_{sym}$.
\begin{equation}
\alpha_{pri}=\frac{4C_{sym}}{T}\left[ \left(\frac{Z_1}{A_1}
\right)^2
  - \left( \frac{Z_2}{A_2}\right)^2\right] =
\frac{4C_{sym}}{T}\left[\left(\Delta \frac{Z}{A}\right)^2\right]
\end{equation}
where $\alpha_{pri}$ is the isoscaling parameter extracted from the
calculated yields of primary fragments, $T$ is the temperature,
$Z_{i}/A_{i}$ is the proton fraction of the initial source with label i. To
extract $C_{sym}$ which is related to symmetry energy ($E_{sym}=
C_{sym}I^{2}$) from data, it is important that the sequential decays
do not affect $\alpha$, $T$ and $[\Delta (Z/A)]^{2}$ significantly.

Figure 8 shows isoscaling plots constructed from final fragments after
sequential decays. Isoscaling is no longer strictly observed over a
large range of isotopes. Furthermore, the distances between elements
are much less regular and the slopes vary from element to element.
The distances between elements are related to the proton isoscaling
parameter, $\beta$. Experimentally, the trends and magnitudes in
both $\alpha$ and $\beta$ are similar \cite{ref19,geraci}. The
irregular spacings between elements from the calculations is
probably caused by the Coulomb treatment in different codes. Part of
the lack of smoothness in the trends could come from the lack of
statistics for primary isotopes with low cross-sections. By
restricting the isoscaling analysis to the same set of isotopes
measured in experiments, about 3 isotopes for each element
\cite{ref19,geraci}, most models show that the effect from
sequential decays on isoscaling is negligible as shown in the left
panel of Fig.~9. The solid points refer to the analysis of the
systems with the same charge, system 1 and 2 in Table 1, while the open
points refer to the analysis of the systems with the same mass, system 1
and 3. In the two MMM calculations (triangles), the final fragments
seem to retain more memory of the source than the other models as
shown in Fig.~6, resulting in the final isoscaling parameters
being larger than the primary isoscaling parameters. By restricting
the number of isotopes for fitting, the problems with statistics
from fragment production may be minimized. On the other hand, such
procedure may hide fundamental problems associated with the
sequential decays.

\begin{figure}
\centering
\resizebox{1.0\columnwidth}{!}{%
\includegraphics{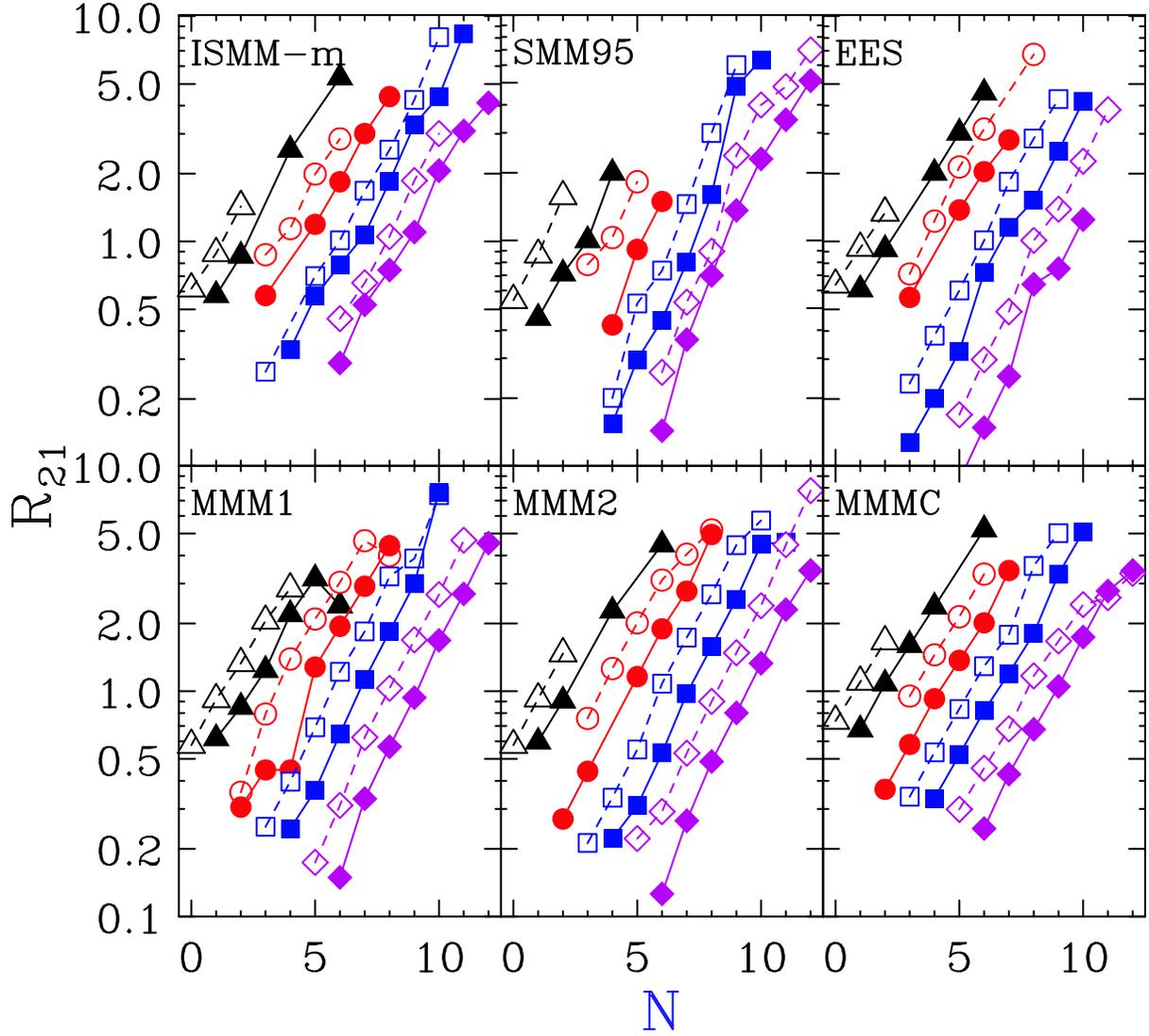}}
\caption{Predicted yield ratios, $R_{21}(N, Z) = Y_{2}(N,
Z)/Y_{1}(N, Z)$ from final fragments. The symbols have the same
convention as in Fig.~7. The lines are drawn to guide the eye. }
\label{fig8}
\end{figure}

All the statistical models except the LGM use the symmetry energy of
stable nuclei in describing the mass of the fragments. Except for
the EES model, the symmetry energy coefficient, $C_{sym}$, remains
constant throughout the reactions. Such prescription may not be
realistic. In a recent study, when different (especially lower)
values of $C_{sym}$ are used in a Markov-chain version of the SMM95
code, the sequential decays effects are very different \cite{ref38}
as shown in the right panel of Fig. 9. The lines denote different
excitation energy (4, 6, 8 MeV) used. For each excitation energy,
calculations have been performed for $C_{sym}=$ 4 (burst symbols), 8
(X symbols), 14 (crosses) and 25 (circular symbols) $MeV$. For
$C_{sym}=25$ MeV, the effect of sequential decays are similar to
those shown in the left panel of Fig.~9. However, for lower
$C_{sym}$ values, the $\alpha$(final) are larger than
$\alpha$(primary). As discussed in \cite{colonnabook}, this trend is
different from those observed in dynamical calculations. A detailed
understanding of the effects of sequential decays on the isoscaling
parameters $\alpha$, the temperature $T$, and the proton fraction
$Z_{i}$/$A_{i}$ \cite{ref39} is necessary before symmetry energy
information can be extracted by applying
 Eq.~(2) to experimental data.
%Without a detailed understanding of the effects of sequential decays on the
% isoscaling parameters $\alpha$, the temperature $T$, and the proton fraction
%$Z_{i}$/$A_{i}$ \cite{ref39}, Eq.~(2) should only be applied to experimental data with caution.

\begin{figure}
\centering
\resizebox{1.0\columnwidth}{!}{%
\includegraphics{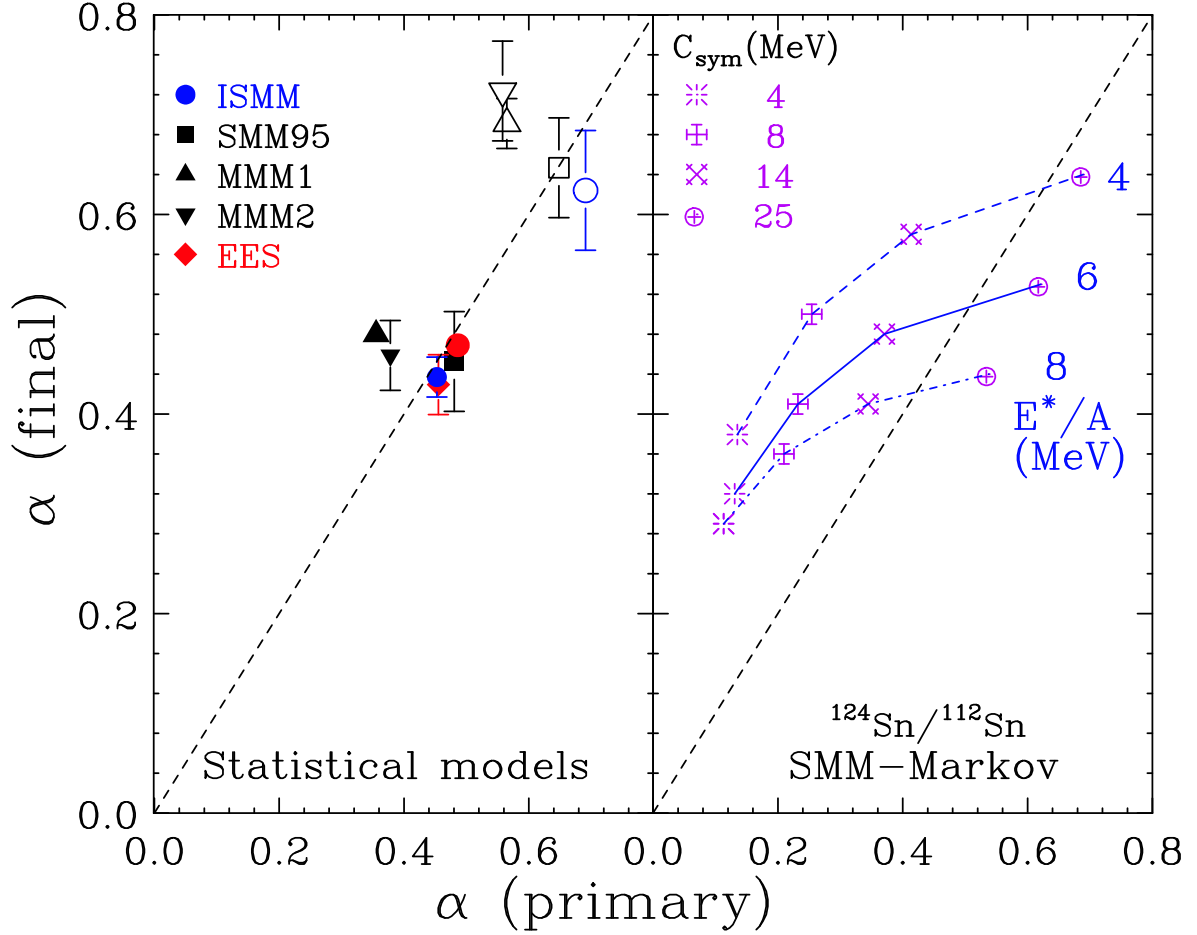}}
\caption{Effect of sequential decays on the isoscaling parameter,
$\alpha$. Left panel shows the results of the statistical models
studied in this work where the $C_{sym}$ in the models assume a
constant value of about 25 MeV and $E*/A=4$ MeV. Right panel shows the
results from the microcanonical version of SMM with Markov-chain
where  $C_{sym}$ varies from 4 (burst symbols), 8 (crosses), 14 (X
symbols) and 25 (circular symbols) MeV and $E*/A$ varies from 4
(dashed line), 6 (solid line) and 8 (dot-dashed line).} \label{fig9}
\end{figure}

%%%%%%%%%%%%%%%%%%%%%%%%%%%%%%%%%%%%%%%%%%%%%%%%%%%%%%%
\subsection{Fluctuations of isotope yield ratio temperatures} %yield ratios}    %% 3.5
%%%%%%%%%%%%%%%%%%%%%%%%%%%%%%%%%%%%%%%%%%%%%%%%%%%%%%%%%%
Ideally, a model should predict isotope cross-sections such as those
shown in the right panels of Fig.~5. All model comparisons involve
the production of primary fragments and their decays. To disentangle
the two parts of the calculations from the final fragments and to
evaluate the sequential decay portion of the calculations, we need
another observable that is mainly sensitive to the structural decay
information, an important ingredient in sequential decay models. It
has been shown that the fluctuations observed in the isotope yield
temperatures are sensitive to the sequential decay information
\cite{ref2,ref40}.

The isotope yield ratio thermometer is defined as \cite{albergo}
\begin{equation}  %% labeled as (2)
\mathrm{T}=\frac{B}{\ln{a\cdot R}}
\end{equation}
where \textbf{$B$} is a binding energy parameter, \textbf{$a$} is
the statistical factor that depends on statistical weights of the
ground state nuclear spins and \textbf{$R$}  is the ground state
isotope yield ratio,
\begin{equation}  %% labeled as (3)
\mathrm{R}=\frac{Y(A_1,Z_1)/Y(A_1+1, Z_1)}{Y(A_2,Z_2)/Y(A_2+1,Z_2)}
\end{equation}

\begin{figure}
\centering
\resizebox{1.0\columnwidth}{!}{%
\includegraphics{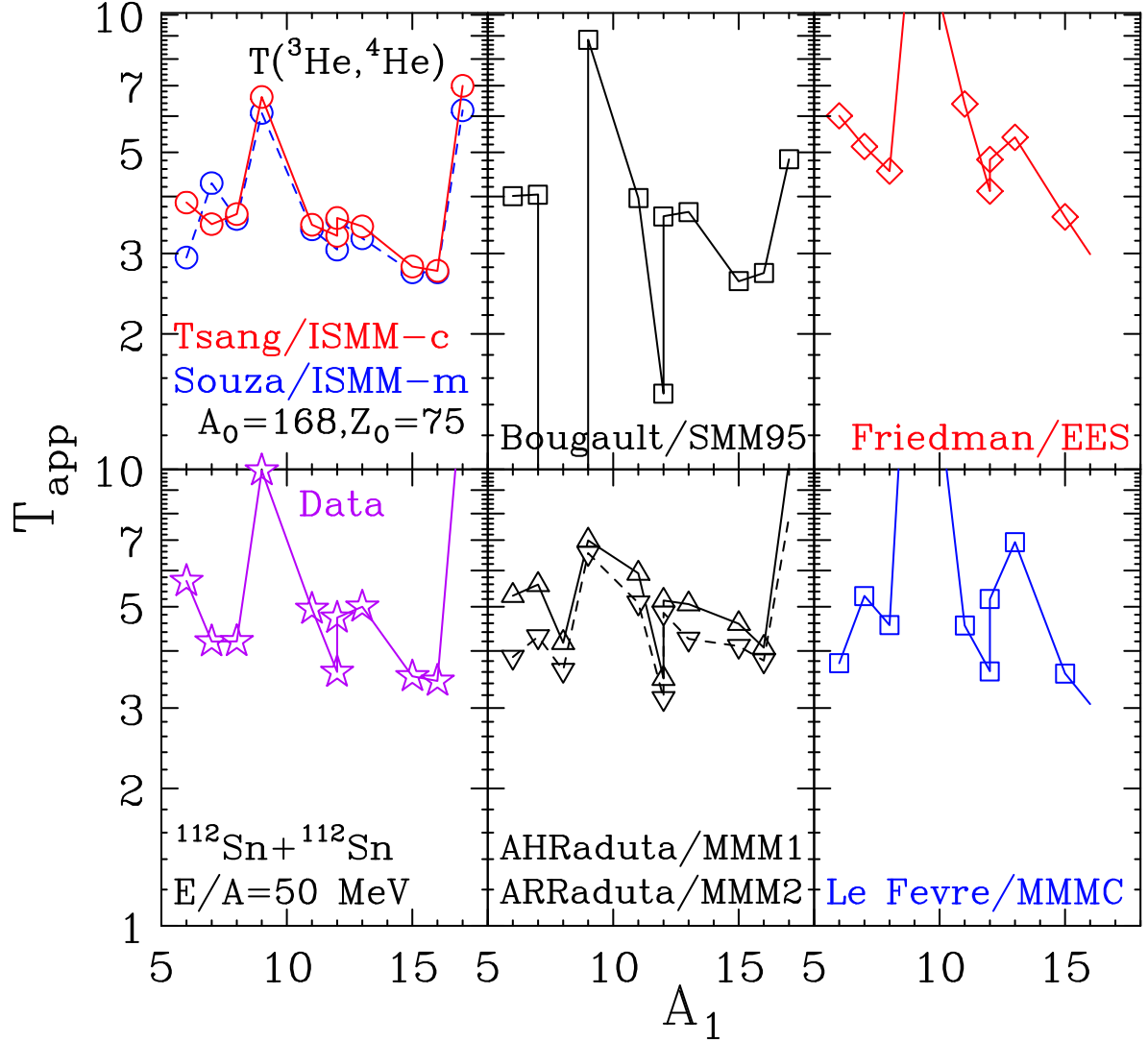}}
\caption{Isotopic temperatures $T(^{3}$He,$^{4}$He) constructed
from different  isotope pairs in the numerator of Eq.~(4) and the
ratios of $Y(^{3}$He)/$Y(^{4}$He) in the denominators are plotted as
a function of $A_1$. The data \cite{ref11b} are plotted in the
bottom left corner for reference. Models with similar decay codes
such as ISMM-c and ISMM-m (top left panel) and MMM1 and MMM2 (bottom
middle panel) are plotted together.} \label{fig10}
\end{figure}

%There is no consensus regarding $T$ defined in Eq.~(3) being the temperatures of the freeze-out source [1]. Such discussion is beyond the scope of this article.
%We will not discuss here the use of $T$ defined in Eq.~(3) for
%determining the temperature of the freeze-out source (see
%contribution of A. Keli{\'c} et al. to this volume \cite{ref31}).
% For lack of better terminology, we call T temperature or isotope yield ratio or thermometer.
In this section, our discussion is mainly focused on using $T$ as a
tool to evaluate the modeling of sequential decays. More details
about using $T$ as the temperature of the freeze-out source can be
found in ref.~\cite{ref31}. It is possible to construct many
different thermometers from various combinations of the isotope
yields using Eqs.~(3) and (4) \cite{ref40}. In the grand-canonical
approximation, if all fragments are produced directly in their ground states,
these temperatures should all have the same value as the temperature
of the initial system. Experimentally, we see large fluctuations of
these isotope yield temperatures \cite{ref40,ref41,ref42}, i.e. $T$
depends on the specific combinations of isotopes used in Eq.~(4).
Without sequential decay corrections, the measured temperature is
not the source temperature. 
%By convention, $T$ is often called the
%apparent temperature, $T_{app}$ as denoted in Fig. 10. 
Because of the fluctuations, the experimental measured temperatures are 
usually called $T_{app}$ as denoted in Fig. 10. 
As an example, we show $T(^{3}$He,$^{4}$He)
constructed with $Y(^{3}$He) and $Y(^{4}$He) yields in the
denominators ($A_{2}$=3, $Z_{2}$=2) but different isotope pairs in
the numerators of Eq.~(4). Specifically, we will examine eleven
$T(^{3}$He,$^{4}$He) thermometers constructed with the yields of the
following isotope pairs in the numerators:
\\$Y(^{6}$Li)/$Y(^{7}$Li),
$Y(^{7}$Li)/$Y(^{8}$Li), $Y(^{8}$Li)/$Y(^{9}$Li),\\
$Y(^{9}$Be)/$Y(^{10}$Be), $Y(^{11}$B)/$Y(^{12}$B),
$Y(^{12}$B)/$Y(^{13}$B),\\$Y(^{12}$C)/$Y(^{13}$C),
$Y(^{13}$C)/$Y(^{14}$C),
$Y(^{15}$N)/$Y(^{16}$N),\\$Y(^{16}$O)/$Y(^{17}$O), and
$Y(^{17}$O)/$Y(^{18}$O). 

These isotope pairs are chosen because the
data for the central collisions  of $^{112}$Sn+$^{112}$Sn at $E/A$=50
MeV are available \cite{ref2,ref18}. $T(^{3}$He,$^{4}$He) are
constructed from the values of $a$ and $B$ listed in ref.~\cite{ref11b}.
These temperatures are plotted as a function of $A_1$
in the lower left panel of Fig.~10. To get a glimpse of how well
different evaporation codes which are coupled to the statistical
multifragmentation models listed in Table 1 simulate sequential
decays, $T(^{3}$He,$^{4}$He) constructed with the final fragments
produced from the different statistical models are plotted in the
remaining panels of Fig.~10.
%One can compare the fluctuations to the fluctuations observed in the central collisions of the Sn isotopes (lower left panel).
Instead of assuming a constant value, $T(^{3}$He,$^{4}$He)
fluctuates in all models. This suggests that decays to low lying
excited states occur. If a significant fraction of the particles
de-excite to the gamma levels below the particle decay thresholds,
the ground state cross-sections are modified. Such contaminations
may have caused the higher temperatures determined from the yields
of $^{9}$Be ($A_{1}=9, Z_{1}=4$) and $^{18}$O ($A_{1}+1=18,
Z_{1}=8)$ which have several
%large numbers of
low lying excited states below the neutron thresholds. Similar
fluctuations have been observed in different reaction systems at
different temperatures \cite{ref11b,ref40,ref41,ref42}. They mainly
originate from the detailed structure of the excited states. Thus
the fluctuations in the isotope temperature provide a sensitive
tool to evaluate whether proper decay levels have been taken into
account in a code.

These fluctuations are mainly determined by the sequential decay
portion of the code. Models with the same decay codes exhibit nearly
the same fluctuations even 
\linebreak[4]
though the primary IMF multiplicities and
mass distributions are different. For example, different freeze-out
assumptions used in the two MMM codes result in very different mean
IMF multiplicities (Fig.~1) and different residue distributions
(Fig.~2). However, the isotope yield ratio temperatures have the
same trends (bottom middle panel) suggesting that sequential decays
mask off some initial differences in the source. The fluctuations in
ISMM-c and ISMM-m are similar (top left panel). Since the MSU-decay
code incorporates the most structural information for the light
fragments ($Z<$15), $T(^{3}$He,$^{4}$He) determined from the two
ISMM codes that employ the MSU-decay as after-burners reproduce the
trend of the experimental fluctuations the best (top left panel).
However, $T(^{3}$He,$^{4}$He) is lower than the input temperature of
4.7 MeV suggesting that the sequential decay effects on the initial
temperature can be substantial. As $^{9}$Li isotopes are not
produced in the SMM95 code, the temperatures involving this isotope
drops (top middle panel). Individual temperature values do not agree
among models even though the initial input to the fragmenting source
is the same. The differences in the isotope yield ratio temperatures
probably reflect the difference in the decay codes.

%%%%%%%%%%%%%%%%%%%%%%%%%%%%%%%%%%%%%%%%%%%%%%%%%%%%%%%%%%%%%%%%%%%%%%%%%%%%%%%%%%%%%%%%%%%%%%%%%%
\section{Evaporation models:}                %% section 4.
%%%%%%%%%%%%%%%%%%%%%%%%%%%%%%%%%%%%%%%%%%%%%%%%%%%%%%%%%%%%%%%%%%%%%%%%%%%%%%%%%%%%%%%%%%%%%%%%%%
Before comparing calculated results with data, all hot fragments
produced in any models must undergo decay. Unfortunately the task to
simulate sequential decays has proved to be rather difficult due to
the lack of complete information on nuclear structures and level
densities. In this section, we compare five sequential decay models
(see Table 1) that have been used in many studies. The bench mark
systems are 1) $A_{0}$=168, $Z_{0}$=75 and 2) $A_{0}$=186,
$Z_{0}$=75. The excitation energy is 2 MeV per nucleon. For brevity,
we only discuss three observables, which illustrate the differences
in the codes.

\begin{figure}
\centering
\resizebox{1.0\columnwidth}{!}{%
\includegraphics{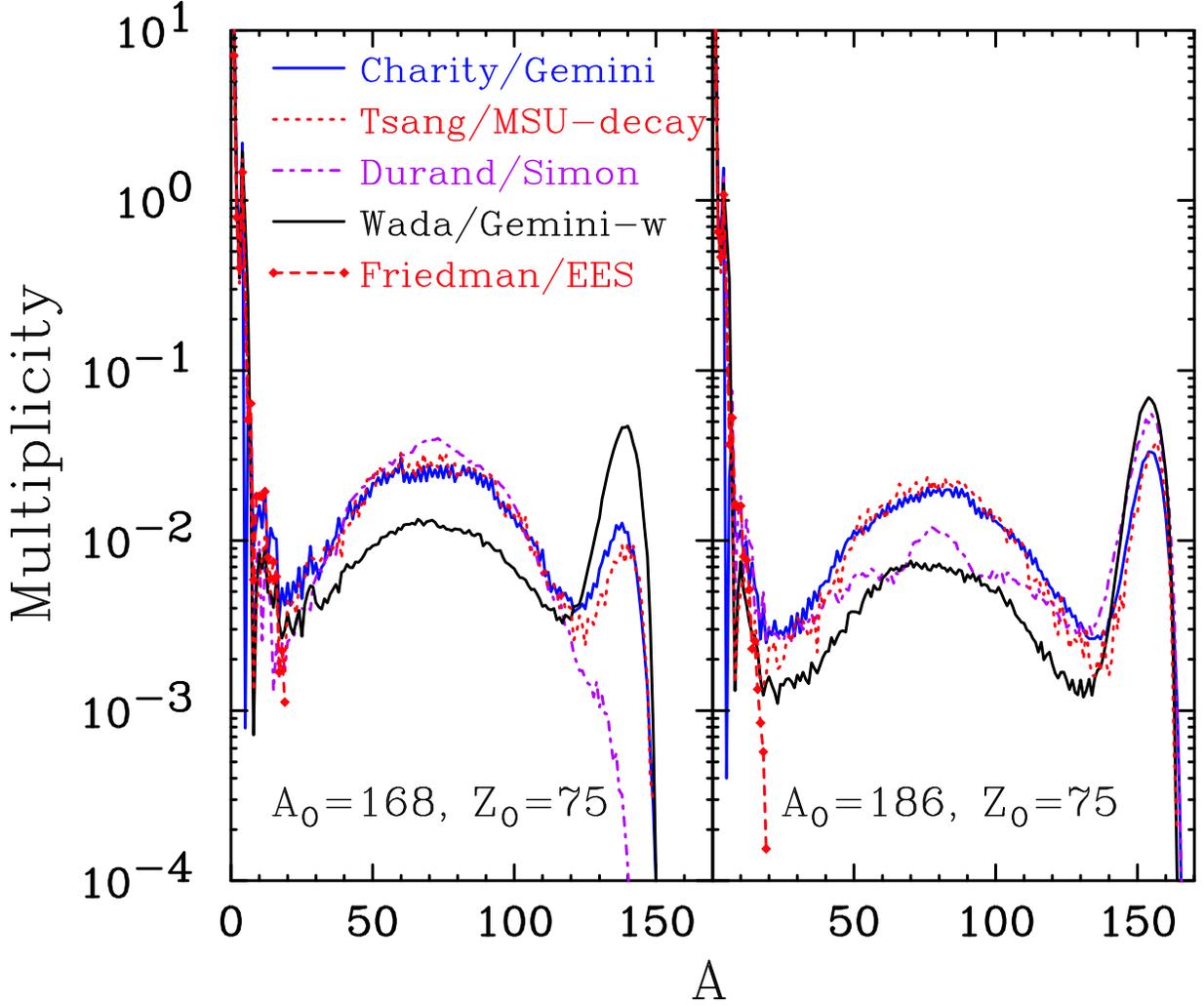}}
\caption{Predicted mass distributions from the five evaporation codes
listed  in Table 1 for the neutron-deficient system 1 (left panel)
and neutron-rich system 2 (right panel).} \label{fig11}
\end{figure}

%%%%%%%%%%%%%%%%%%%%%%%%%%%%%%%
\subsection{Mass distributions}   %% 4.1
%%%%%%%%%%%%%%%%%%%%%%%%%%%%%%%%%
Figure 11 shows the mass distributions from the decay of the
$A_{0}$=168, $Z_{0}$=75 system (left panel) and $A_{0}$=186,
$Z_{0}$=75 system (right panel). Contrary to the near exponential
decrease of the production of fragments with increasing mass in
multifragmentation processes (Fig.~2), most evaporation models
de-excite by emitting LCPs, leaving a residue. Fission is also a
significant de-excitation mode in this mass region, resulting in a
hump at about 10 mass units less than $A_{0}$/2. The inability of the
EES model (symbols joined by dashed lines) to track fragments larger
than A=20 results in the artificial truncation of the mass
distribution. Since the MSU-decay uses Gemini to decay fragments
with $Z>$15, results from Gemini (solid line) and MSU-decay models
(dotted line) are very similar. In principle, Gemini-w (dashed line)
should be the same as Gemini. However, an older version of Gemini
was incorporated and Germini-w gives much larger residue
cross-sections and correspondingly smaller fission fragment and IMF
cross-sections. The event generator code, SIMON (dot-dashed line)
has very different mass distributions than the other codes, e.g. it
does not produce residues in the $A_{0}$=168, $Z_{0}$=75 system
(left panel).

%%%%%%%%%%%%%%%%%%%%%%%%%%%%%%%%%%%
\subsection{Isoscaling}    % 4.2
%%%%%%%%%%%%%%%%%%%%%%%%%%%%%%%%%%%%%

Primary fragments produced from nearly all statistical
multifragmentation codes observe isocaling, rigorously.
\linebreak[4]
However, isoscaling is not well observed over a large range of
secondary fragments. For this reason, we limit the number of
isotopes to three for each element, similar to those measured in
experimental data. We use this observable to examine the differences
between different models in Fig.~12. The symbol convention of
Figs.~7 and 8 is used, i.e. symbols are yield ratios and
lines are best fits. Isoscaling is reasonably reproduced except for
SIMON. For the MSU-dacay and EES (not shown) decays, the results are
similar to those of ISMM and EES calculations shown in Fig.~8.
Except for $^{6}$He yield ratios, Gemini exhibits very good
isoscaling. The isoscaling from Gemini-w fragments is not as good.
The same problems that cause SIMON to produce different mass
distributions could be the cause for the non-observation of
isoscaling behavior.

%%%%%%%%%%%%%%%%%%%%%%%%%%%%%%%%%%%%%%%%%%%%%%%%%
\subsection{Fluctuation of isotope yield ratio temperatures}   % 4.3
%%%%%%%%%%%%%%%%%%%%%%%%%%%%%%%%%%%%%%%%%%%%%%%

\begin{figure}
\centering
\resizebox{1.0\columnwidth}{!}{%
\includegraphics{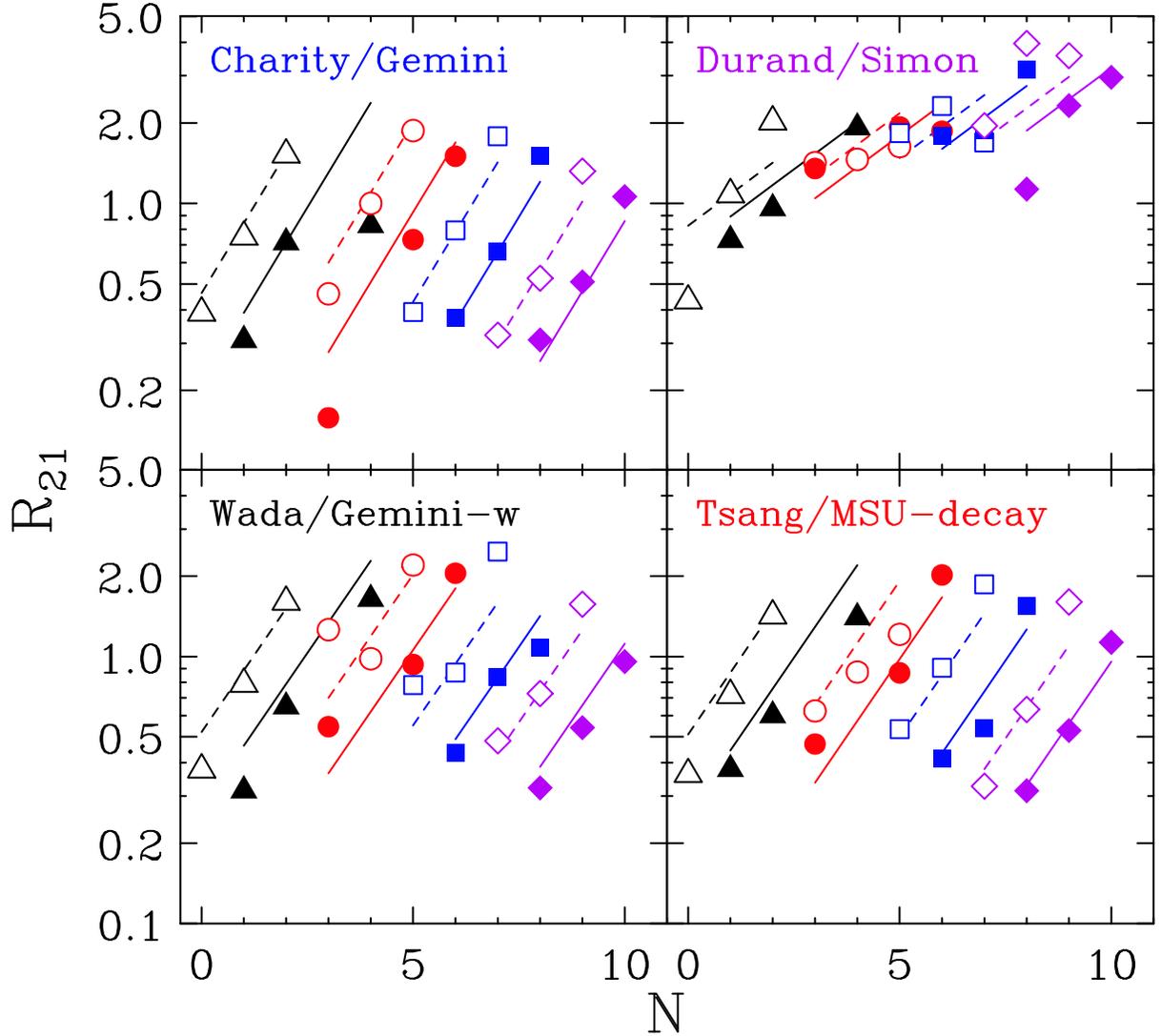}}
\caption{Predicted ratios, $R_{21}(N,Z) = Y_{2}(N,Z)/Y_{1}(N,Z)$
of fragments evaporated from system 2 and system 1.
Each panel presents the results from one model calculation. The
results from the EES model are not plotted here as they are similar to those
shown in Fig.~8. } \label{fig12}
\end{figure}

In Fig. 13, we show T($^{3}$He, $^{4}$He) constructed with
Y($^{3}$He)  and Y($^{4}$He) yields in the denominators ($A_{2}=3,
Z_{2}$=2) but different isotope pairs in the numerators of Eq.~(4)
as discussed in Section 3.5. For reference, the Sn data are plotted
in the lower left panel of Fig.~13 as a function of $A_{1}$. As
light-particle structure information has been included in EES,
Gemini, and MSU-decay codes, they reproduce the fluctuations
observed experimentally rather well as shown in the top three panels
in Fig.~13. On the other hand, SIMON and Gemini-w do not reproduce
the general trends suggesting that the sequential decays are not
properly taken into account in these codes.

Most of the isotope yield ratio temperatures from EES, Gemini and
MSU-decay  calculations are below 4 MeV, the input temperature of
the source. Sequential decay effects are expected to reduce the
initial temperature. It is interesting to note that of the three
models that reproduce the fluctuations, the average temperature is
the highest for the EES model and lowest for the MSU-decay model.
This can be explained by the amount of structural information
included in individual models. EES incorporates only a few low lying
excited states while the MSU-decay model incorporates most of the
experimental level information for nuclei with Z$<$15. The
availability of a large number of decay levels in the latter code
reduces the ground-state cross-sections more than in other
calculations. This suggests that even at low excitation energy (2
MeV), sequential decays still significantly affect isotope yields.

\begin{figure}
\centering
\resizebox{1.0\columnwidth}{!}{%
\includegraphics{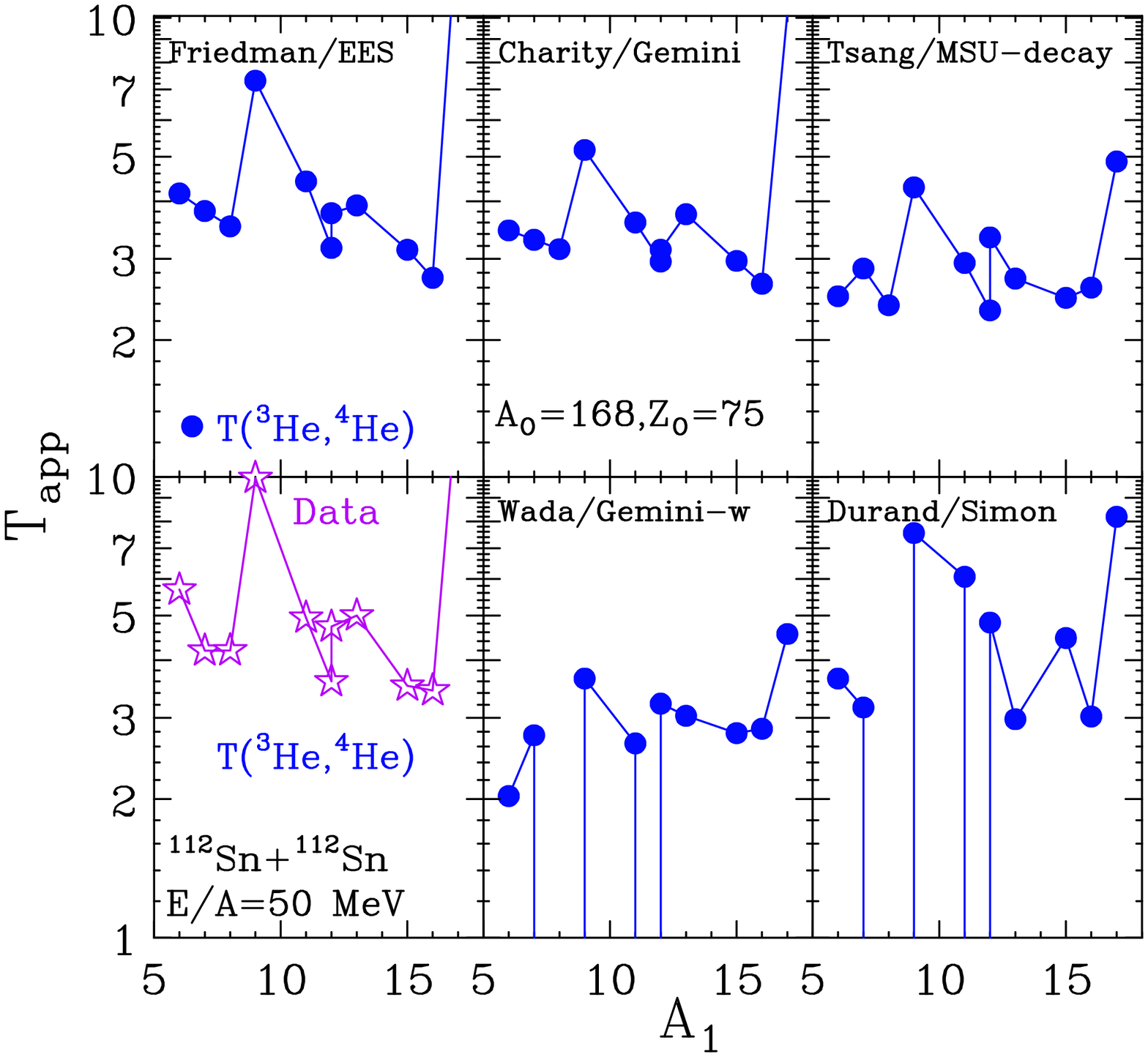}}
\caption{Isotopic temperature $T(^{3}$He,$^{4}$He) plotted as a
function of $A_{1}$. For reference, the data \cite{ref11b} are
plotted in the bottom left corner. }
\label{fig13}
\end{figure}

%%%%%%%%%%%%%%%%%%%%%%%%%%%%%%%%%%%%%%
\section{Summary and Conclusions}
%%%%%%%%%%%%%%%%%%%%%%%%%%%%%%%%%%%%%%%%%%
In summary, we have made comparisons of experimental observables
using ten statistical multifragmentation codes. The general trends
are similar among models suggesting that these models can provide
important physical insights for the primary fragments and
multifragmentation process. However, details in any single
observable differ between models. The largest differences are
observed in raw observables such as individual isotope yields, mass
and charge distributions while the mean values of an observable such
as IMF multiplicity, the mean fragment asymmetry $<N/Z>$ or mean
mass $<A>$ of an element do not show as large differences. The
effects of sequential decays on isoscaling parameters are not well
understood.

As sequential decay codes are important to both dynamical and
statistical  models, we also compare five widely used codes.
Relatively accurate structural information and experimental masses
are required in evaporation models to reproduce the fluctuations of
isotope yield temperatures. Such sensitivity allows one to evaluate
the sequential decay properties of the evaporation codes.

The observables studied here are by no means an exhaustive list.
However, these observables, which can be constructed easily from the
isotope yields, provide important bench-marks to test any
multifragmentation models or evaporation codes that describe
sequential decays.

%%%%%%%%%%%%%%%%%%%%%%%%%%%%%%%%%%%%%%
\section{Acknowledgment}
%%%%%%%%%%%%%%%%%%%%%%%%%%%%%%%%%%%%%%%%%%
The authors wish to thank Dr. C. Bertulani for his help in the
beginning of this project. MBT acknowledges the support of the
National Science Foundation under Grant No. PHY-01-10253.

\end{document}